# Experimental evidence of (M1+E2) mixed character of the 9.2 keV transition in $^{227}$Th and its consequence for spin-interpretation of low-lying levels

A. Kovalík [a,b], A.Kh. Inoyatov [a,c], L.L. Perevoshchikov [a], M. Ryšavý [b], D.V. Filosofov [a], P. Alexa [d], J. Kvasil [e]

[a] *Dzhelepov Laboratory of Nuclear Problems, JINR, 141980 Dubna, Moscow Region, Russian Federation*
[b] *Nuclear Physics Institute of the ASCR, CZ-25068 Řež near Prague, Czech Republic*
[c] *Institute of Applied Physics, National University, University Str. 4, 100174 Tashkent, Republic of Uzbekistan*
[d] *Department of Physics, VSB-Technical University of Ostrava, 17. listopadu 2172/15, 708 00 Ostrava, Czech Republic*
[e] *Institute of Particle and Nuclear Physics, Charles University, CZ-18000, Praha 8, Czech Republic*

The 9.2 keV nuclear transition in $^{227}$Th populated in the β-decay of $^{227}$Ac was studied by means of the internal conversion electron spectroscopy. Its multipolarity was proved to be of mixed character M1+E2 and the spectroscopic admixture parameter $\delta^2(E2/M1)=0.695\pm0.248$ ($|\delta(E2/M1)|=0.834\pm0.210$) was determined. Nonzero value of $\delta(E2/M1)$ raises a question about the existing theoretical interpretation of low-lying levels of $^{227}$Th.

## 1. Introduction

The interpretation of the level structure of $^{227}$Th is still a major problem mainly due to the lack of experimental information on the low-lying levels of $^{227}$Th and a long-standing controversy [1] about the spin-parity of the $^{227}$Th ground state, which represents the basis for all other level spin-parity assignments. Until recently only two excited levels at 9.3 and 24.5 keV were known from the β$^-$-decay of $^{227}$Ac [2] and an additional one at 77.7 keV from the α-decay study of $^{231}$U [3] feeding the former two.

In earlier compilations [4,5], the spin-parity 3/2$^+$ was suggested for the $^{227}$Th ground state. In subsequent compilations (including the most recent ones [6,7]), the spin-parity 1/2$^+$ was recommended for it. Nevertheless, this assignment contradicted the observed distribution anisotropies for the gamma rays following the α decay of $^{231}$U in the α-γ nuclear orientation experiment [8] in which the 3/2$^+$ spin-parity for the $^{227}$Th ground state was assumed. However, later the authors [9] came to a conclusion that the $^{227}$Th ground-state spin is consistent with 1/2 (though their data did not strictly exclude spin≠1/2). This was done on the basis of the measurements [3,10] and results of their investigation [9] of the α-decay of $^{231}$U and the angular distribution of α particles in the decay of low-temperature oriented $^{227}$Th. They deduced that the above-mentioned nuclear-orientation experiment [8] was in error and thus the only experimental evidence for a $^{227}$Th ground-state spin≠1/2 was eliminated.

The present spin-parity assignments of the $^{227}$Th levels are based on the assumption of spin-parity 1/2$^+$, 5/2$^+$ and 3/2$^+$ for the ground state, 9.3 keV and 24.5 keV levels, respectively. This interpretation supposes the pure E2 multipolarity for the 9.3 keV gamma-ray transition (depopulating the lowest excited state of $^{227}$Th). Such a multipolarity was determined as the "most likely" one in Ref. [2] from the conversion-electron data in the study of the β$^-$ decay of $^{227}$Ac but the authors [2] did not excluded *a considerable admixture of the M1 multipolarity*. The above spin-parity assignments were first proposed in Ref. [11] and are also assumed in the most recent nuclear data compilation [7].

Thus, it is obvious that the adopted interpretation [7] of the lowest $^{227}$Th levels is not yet strictly established by experiment. It needs additional experimental data, in particular, reliable and accurate multipolarity determination of the 9.3 keV transition on which the whole spin sequence and band structure above the ground state depends. Therefore, we performed a new study of the low-energy electron spectrum generated in the β$^-$ decay of $^{227}$Ac (in which the ground state and the lowest three excited levels of $^{227}$Th are populated [7], see Fig. 1) using the internal conversion electron spectroscopy (ICES). Results obtained on the 15.1 keV (M1+E2) transition in $^{227}$Th were already published in Ref. [12]. In this work, results on the multipolarity determination of the 9.3 keV transition depopulating the 9.30(3) keV 5/2$^+$ (assumed) level of $^{227}$Th [7] are given.

## 2. Experiment and analysis of the spectra.

The $^{227}$Ac source for the investigation was produced by a sorption of slightly soluble forms of actinium (AcF$_3$) on a carbon polycrystalline foil and its activity was 690 kBq just after the preparation (for details see Ref. [12]).

The electron spectra were measured in sweeps at the 14, 21, and 35 eV instrumental energy resolution with the 2, 3, 5, 6, 8, and 10 eV scanning step by a combined electrostatic electron spectrometer [13,14]. Examples of the measured spectra are shown in Figs. 2, 3.

To decompose the measured conversion electron spectra into the individual components, the approach and the computer code SOFIE (see, e.g., Ref. [15]) was applied. In this approach, the spectral line profile is expressed by a convolution of the Lorentzian (describing the energy distribution of the investigated electrons leaving atoms) with an artificially created function based on the Gaussian. The aim of the latter function is to describe both the response of the spectrometer to the monoenergetic electrons and the observed deformation of the measured electron lines on their low-energy slopes caused by inelastically scattered electrons in the source material. The Monte Carlo procedure is, therefore, involved in the code.

## 3. Transition multipolarity determination.

It should be noted that if one accepts arguments of the most recent nuclear data compilation [7] that the 9.3 keV transition in $^{227}$Th depopulates the first excited state 5/2$^+$ to the ground state 1/2$^+$, then any mixture containing the M1 multipolarity is excluded but the M3 multipolarity is possible in principle (e.g. E2+M3 multipolarity mixture).

For the multipolarity determination, thirteen independent experimental values (i.e., obtained from different measurements) of the M- and N-subshell conversion line ratios were used: $M_1/M_2$ = 0.031(11), 0.027(9); $M_1/M_3$ = 0.025(9), 0.023(7); $M_2/M_3$=0.852(10); $M_4/M_3$ = 0.019(6), 0.021(4); $M_5/M_3$ = 0.019(4), $N_1/N_3$ = 0.02(3), 0.012(13); $N_2/N_3$ = 0.84(7), 0.84(2); and $N_5/N_3$ = 0.044(17).

The theoretical internal conversion coefficients for the M1, M3, and E2 multipolarities and for transition energy of 9245 eV [16] were calculated employing the computer code NICC [17] using the potential [18] for a neutral thorium atom and the thorium electron binding energies [19]. They are presented in Table 1 (marked as NICC). In order to minimize a possible influence of the theoretical ICC evaluation approach on the multipolarity determination, we applied also another (widely used) method BrICC [20] using the internet calculator [21]. These ICC's are also presented in Table 1 (marked as BrICC).

The transition multipolarity was then determined using the program CFIT [22] which fits the theoretical ICC's and/or ICC ratios into the experimental values by means of the least-squares method.

As can be seen from Table 1, the $s_{1/2}$ to $p_{1/2}$ subshell ratio values of theoretical M1-ICC are around 8 (for the both ICC sets), while our corresponding experimental ratios do not exceed 0.03 (see our experimental data above). This clearly proves that the 9.2 keV transition cannot be of pure M1 multipolarity.

Then we tested (only with the use of the NICC set) a possibility of a multipolarity mixture of E2+M3. In such a case the ICC $\alpha_i$ is given by the formula $\alpha_i=(1-\Delta)\alpha_i(M3)+\Delta\alpha_i(E2)$, where the subscript $i$ marks the atomic (sub-)shell and $\Delta$ is the admixture of the E2 multipolarity (connected with the $\delta^2(E2/M3)$ admixture parameter commonly used in the gamma ray spectroscopy by the relation $\Delta=\delta^2/(1+\delta^2)$). We obtained $\Delta=1.0\pm(1.0\times10^{-6})$ and the $\chi_\nu^2$ value the same as for the pure E2 multipolarity (see Table 2).

Finally we performed analysis with both sets of theoretical ICC's for the assumed multipolarities, i.e. the pure E2 and the mixture M1+E2 (in the latter case the theoretical ICC is expressed as $\alpha_i=(1-\Delta)\alpha_i(M1)+\Delta\alpha_i(E2)$). The results obtained are presented in Table 2. As can be seen from the table, the received $\chi_\nu^2$ values indicate that the NICC set of the theoretical ICC's is in better agreement with the experiment than the BrICC one. But what is more important, the difference between the $\chi_\nu^2 = 0.89$ for the mixed multipolarity M1+E2 and $\chi_\nu^2 = 1.18$ for the pure E2 in the case of the NICC set (and correspondingly $\chi_\nu^2=1.03$ and 1.35 for the BrICC set) strongly indicates that the 9.2 keV transition is of the mixed multipolarity. To prove such a statement, we applied the statistical method presented in Ref. [23].

The essence of this method is as follows: We have a set of experimental values, $F = (F_1\pm\sigma_1, F_2\pm\sigma_2, ..., F_n\pm\sigma_n)$. We have also two hypotheses explaining the experiment, $f_1$ and $f_2$, which give the corresponding theoretical values, $f_1 = (f_{11}, f_{12}, ..., f_{1n})$ and analogically $f_2$. We should decide which one is true, i.e. which one can be rejected. Let us assume $f_2$ be null hypothesis and we seek the probability that – rejecting it – we reject the true hypothesis. The method consists of study of the expression $M_2 - M_1$ where $M_i=\sum_{j=1}^{n}\frac{(F_j-f_{ij})^2}{\sigma_j^2}$ as a function of independent statistical variables $F_j$.

The result is, that the probability $P$ to reject the true hypothesis (i.e. $f_2$) is $P \leq \frac{1}{2}\left[1-Erf\left(\sqrt{\frac{\eta_{exp}}{2}}\right)\right]$. Here $\eta_{exp} = M_2 - M_1$ evaluated with the experimental values $F$, i.e. a difference of the $\chi^2$'s *not* normalized for one degree of freedom, and *Erf* is the error function, $Erf(x)=\frac{2}{\sqrt{\pi}}\int_0^x \exp(-t^2)dt$.

In our case, $f_1$ is the hypothesis "mixed multipolarities" and $f_2$ is "pure E2". Then $\eta_{exp} = 15.34-10.68 = 4.66$ (for the NICC set). The probability, that rejecting "pure E2" we reject the true hypothesis is $P \leq 0.0155$. This means that there is less than one and half percent probability that the 9.2 keV transition in [227]Th might be of pure E2 multipolarity. We thus conclude that this transition is of the mixed multipolarity of M1 with the admixture $\Delta=0.41\pm0.12$ (i.e., $\delta^2(E2/M1) = 0.695\pm0.248$; $|\delta(E2/M1)|=0.834\pm0.210$) of E2.

## 4. Calculations and Interpretation

[227]Th belongs to the Ra-Th isotopic region where octupole correlations are very important for theoretical description of low-lying excitations (see e.g. [27],[28],[29]). From theoretical point of view there are two types of approaches for treating the stronger octupole correlations in this region.

In the first one it is supposed that octupole correlations of single-nucleon states near the Fermi level are sufficiently strong for the formation of stable octupole deformation (e.g. [27],[28], [29]) and the smaller part of the residual octupole correlations responsible for shape-vibrations around the reflection asymmetric shape) is usually neglected.

In the second one the nuclear mean field is reflection symmetric but strong octupole correlations in the residual interaction leads to relatively large collective vibrational components in eigenstates of the intrinsic Hamiltonian (see e.g. [10],[30] and citations therein).

In order to describe experimental spectrum and corresponding transition rates it is necessary to take into account rotational degrees of freedom usually done in the framework of the particle-rotor model. In the case of odd-A nuclei the intrinsic (single-particle and vibrational) degrees of freedom are mixed with rotational ones by means of the Coriolis (ΔK=1) interaction between rotational bands. This mixing is very important for the interpretation of the experimental odd-A nucleus spectrum (see [10],[29],[30]). As concerns nucleus $^{227}$Th, the interpretation in Ref. [10] was based on the calculation of the strongly Coriolis mixed rotational bands built on the lowest K=1/2$^+$, K=3/2$^+$ and K=5/2$^+$ intrinsic reflection symmetric states involving octupole vibrational components. This strong Coriolis mixing together with relatively high positive decoupling parameter for the ground K=1/2$^+$ rotational band lead to the interpretation of the lowest positive parity levels 0.0 keV, 9.2 keV, 77.7 keV and 88.0 keV to be members of the ground rotational band with the non-standard sequence of spins I$^\pi$ = 1/2$^+$, 5/2$^+$, 3/2$^+$, and 9/2$^+$, respectively, and levels 24.3 keV and 127.3 keV being 3/2$^+$- and 5/2$^+$- members of rotational band built on the lowest K=3/2$^+$ intrinsic state. This interpretation was also in agreement with the former reflection asymmetric mean field calculations in [27] where the presence of mutually close neutron orbitals with K$^\pi$= 1/2$^+$ and 3/2$^+$ near the neutron Fermi energy was found. However, there are no calculations of Coriolis mixing rotational bands based on the reflection asymmetric mean field in [27] and in the literature at all.

As it was mentioned in the Introduction from experimental point of view the spin-parity ascription 1/2$^+$ to the $^{227}$Th ground state, I$^\pi$ = 5/2$^+$ for 9.2 keV level and I$^\pi$ = 3/2$^+$ for 24.3 keV level cannot be considered as unambiguous. Nevertheless, from systematic investigation of low-spin spectra of odd-A nuclei from Ra-Th region in the connection with the intrinsic single-neutron scheme near the Fermi level (see e.g. [10] and citations therein) one can expect that at least one of the levels 0.0 keV, 9.2 keV, 24.3 keV corresponds to I$^\pi$ = 1/2$^+$. If it is so then the nonzero mixing δ(E2/M1) obtained in this paper for the transition 9.2 keV → 0.0 keV and nonzero δ(E2/M1) parameters for transitions 24.3 keV → 9.2 keV (see [12]) and 24.3 keV → 0.0 keV (see [16]) imply the conclusion that the ascription 5/2$^+$ for any of these levels (0.0 keV, 9.2 keV and 24.3 keV) is excluded. By other word it means that the presence of level with 5/2$^+$ in the lowest part of the $^{227}$Th spectrum is in contradiction with the measured mixing ratios δ(E2/M1) for transitions mentioned above.

In view of these facts, we tried to modify the Coriolis mixing calculations performed in [10] with the aim to shift the 5/2$^+$- level above 50 keV in order the levels 0.0 keV, 9.2 keV and 24.3 keV correspond to 1/2$^+$ or 3/2$^+$ and in such a way to be consistent with observed mixed E2+M1 transitions among them. Simultaneously we want to keep reasonable agreement between experimental and theoretical energies of higher lying levels in $^{227}$Th similarly as was done in [10]. We used very similar approach as in [10], that means the Quasiparticle-Phonon Model (QPM) of the intrinsic Hamiltonian with the reflection symmetric Nilsson mean field, monopole pairing interaction and long-range quadrupole-quadrupole and octupole-octupole residual interaction. Rotational degrees of freedom were described within the standard axially symmetric rotor model with Coriolis coupling involved. Detailed description of the model can be found in [26].

In order to obtain more realistic values of deformation parameters $\varepsilon_2$ and $\varepsilon_4$ than those used in [10] the equilibrium deformation of the $^{226}$Th (even-even core of $^{227}$Th) was searched for by the minimizing the Skyrme-BCS total mean field energy with the SV-bas Skyrme interaction parametrization (see [24]) for two situations:

(i) Only deformation parameters $\varepsilon_2$ and $\varepsilon_4$ were allowed to be changed. In this case reflection symmetric equilibrium deformation $\varepsilon_2 = 0.186$, $\varepsilon_2 = -0.15$ was found.

(ii) Deformation parameters $\varepsilon_2$, $\varepsilon_3$ and $\varepsilon_4$ were simultaneously varied. Then the reflection asymmetric equilibrium values $\varepsilon_2 = 0.17$, $\varepsilon_3 = 0.14$, $\varepsilon_4 = -0.14$ were obtained.

Equilibrium values of the total mean field energy for these two cases were practically the same. This means that it is not possible to conclude if the mean field is reflection symmetric or asymmetric.

In the next step the eigen problem of the intrinsic QPM Hamiltonian was solved with deformation parameters from the case (i) above. The lowest five eigenstates of the intrinsic Hamiltonian are listed in the Table 3. One can see that all these eigen states have octupole vibrational components.

In the last step the total Hamiltonian in the laboratory frame involving the standard axially-symmetric rotor with Coriolis coupling and fixed inverse moment of inertia, $\hbar^2/2J = 10$ keV for all rotational bands was diagonalized (see [26] for details). In the calculation of electromagnetic transitions fixed effective charges (taking into account CoM motion (see [25],[26]) and the value of the intrinsic quadrupole moment $Q_0 = 821$ e fm$^2$ calculated from the known $^{226}$Th B(E2,$2_1^{+} \to 0_1^{+}$) core transition) and the rotation gyromagnetic factor $g_R = Z/A$ were chosen.

We then investigated the effect of different Coriolis and recoil attenuation factors, $\eta_{cor}$ and $\eta_{rec}$, on the spin and parity sequences of the lowest states in $^{227}$Th in the laboratory frame. The recoil interaction pushes (K=3/2$^-$) up and the fourth excited intrinsic state (K=1/2$^+$) down. For $\eta_{rec} < 0.12$ we get the 3/2$^-$ ground state, for $0.12 < \eta_{rec} < 0.64$ the 3/2$^+$ ground state (the first excited intrinsic state) and for $\eta_{rec} > 0.64$ the 1/2$^+$ ground state. The sequence of the lowest three states is controlled by $\eta_{cor}$. For lower values of $\eta_{cor}$ there exist two regions with sequences 3/2$^+$, 1/2$^+$, and 3/2$^+$ ($\eta_{rec} \sim 0.4 - 0.63$, $\eta_{cor} < 0.55$) and 1/2$^+$, 3/2$^+$, and 3/2$^+$ ($\eta_{rec} > 0.63$, $\eta_{cor} < 0.55 - 0.6$). For higher values of $\eta_{cor}$ one of the two 3/2$^+$ states is replaced by a 5/2$^+$ state. The obtained model energies slightly favor the first sequence because for higher values of $\eta_{rec}$ the lowest 3/2$^-$ state lies too high in energy that is in contradiction with the experimental value of 37.9 keV. However, the theoretical values of the admixture parameter $\delta^2$(E2/M1) fit better the second sequence (see Table 4).

4. Conclusion

Highly efficient low-energy nuclear electron spectroscopy technique developed in the JINR Dubna was applied for the analysis of the 9.2 keV transition in $^{227}$Th populated in the β$^-$ decay of $^{227}$Ac. The mixed multipolarity M1+E2 proved for this transition in the present work together with the same multipolarity character determined experimentally in Refs. [12] and [16] for the 15.1 keV and 24.3 keV transitions in $^{227}$Th, respectively, cause doubts for existing spin assignments 1/2$^+$, 5/2$^+$ and 3/2$^+$ for the lowest levels 0.0 keV, 9.2 keV and 24.3 keV, respectively, of $^{227}$Th (see [10]). Particularly, the assignment I$^\pi$ = 5/2$^+$ for any of these levels is excluded in such a case.

Up to date interpretation of low-lying spectrum of $^{227}$Th is influenced by strong Coriolis coupling of rotational bands built on the lowest intrinsic K$^\pi$ = 1/2$^+$ or K$^\pi$ = 3/2$^+$ states which pulls down the I$^\pi$ = 5/2$^+$ state (see [10]). We tried to play with parameters of the rotor + QPM Hamiltonian (the same Hamiltonian as in [10]) with the aim to shift the I$^\pi$ = 5/2$^+$ level up in the spectrum and simultaneously to keep relatively good agreement of the calculated and experimental spectra also for higher-lying levels. Particularly, we varied the strength of the Coriolis coupling and found that for a reduced strength by the attenuation factor of $\eta_{cor} \sim 0.6$ the three lowest states (1/2$^+$, 3/2$^+$, and 3/2$^+$) are compatible with the present experimental data but we failed to reproduce the position of the lowest negative parity states.

In order to prove the new interpretation of the current experimental data it is necessary to use more precise theoretical approaches, e.g., an approach with the laboratory Hamiltonian involving Coriolis mixing of rotational bands based on reflection asymmetric intrinsic states (which has not been used in practice up to now) and that could solve the problem of the position of the lowest negative parity states. From experimental point of view more information about low-energy transitions connecting low-lying levels is also desirable (for instance data about E1 transitions between low-lying levels of opposite parity rotational bands with given K because such strong E1 transitions are indications of stable octupole deformation).

**Acknowledgement**

We highly appreciate Dr. Balraj Singh (McMaster Univ., Hamilton, Canada) for initiating of this investigation, permanent support and valuable discussions.

This work was partly supported by Project founded by the MEYS of the Czech Republic under the contract LTT18021 and by projects founded by the Czech Science Agency (Project No. 19-14048S).

**Table 1**
Theoretical subshell internal conversion coefficients for the M1, M3 and E2 multipolarities calculated in the present work for the 9.245 keV [16] transition in $^{227}$Th using both the computer code NICC [17] with the potential [18] (for a neutral thorium atom) and the BrICC approach [20,21].

| Atomic subshell | M1 | | M3 | E2 | |
|---|---|---|---|---|---|
| | NICC | BrICC | NICC | NICC | BrICC |
| $M_1$ | 7.7901(+2)[a] | 6.84(+2) | 8.4945(+7) | 2.0144(+3) | 1.91(+3) |
| $M_2$ | 9.2663(+1) | 8.54(+1) | 1.0884(+6) | 1.1840(+5) | 1.119(+5) |
| $M_3$ | 5.8425(+0) | 5.66(+0) | 1.0519(+9) | 1.4010(+5) | 1.340(+5) |
| $M_4$ | 1.0925(+0) | 1.054(+0) | 8.7698(+6) | 2.2461(+3) | 2.14(+3) |
| $M_5$ | 5.5793(-1) | 5.41(-1) | 3.2546(+7) | 1.6545(+3) | 1.604(+3) |
| | | | | | |
| $N_1$ | 2.1201(+2) | 1.83(+2) | 2.9510(+7) | 6.1683(+2) | 5.69(+2) |
| $N_2$ | 2.4937(+1) | 2.26(+1) | 5.4154(+5) | 3.1899(+4) | 2.98(+4) |
| $N_3$ | 1.5402(+0) | 1.490(+0) | 2.9760(+8) | 3.7505(+4) | 3.58(+4) |
| $N_4$ | 2.9294(-1) | 2.82(-1) | 2.8823(+6) | 6.2140(+2) | 5.88(+2) |
| $N_5$ | 1.4749(-1) | 1.427(-1) | 1.0026(+7) | 4.2783(+2) | 4.13(+2) |

[a] 7.7901(+2) means 7.7901x10$^2$

**Table 2**
The results of the transition multipolarity determination for the 9.2 keV transition in $^{227}$Th obtained in the present work using the computer code [22] for two different sets of ICC's, namely NICC [17,18] and BrICC [20,21].

| ICC calculations | M1 + E2 | | | E2 | |
|---|---|---|---|---|---|
| | Δ | $\chi_\nu^{2}$ [b] | $\nu$ [a] | $\chi_\nu^{2}$ [b] | $\nu$ [a] |
| NICC | 0.41±0.12 | 0.89 | 12 | 1.18 | 13 |
| BrICC | 0.38±0.11 | 1.03 | 12 | 1.35 | 13 |

[a] $\nu$ is the number of degrees of freedom.
[b] $\chi_\nu^2$ is normalized $\chi^2$ per one degree of freedom.

## Table 3

Energy and structure (single-quasiparticle and quasiparticle + even-even core vibrational phonon components) of the lowest five intrinsic excitations (eigenstates of the intrinsic Hamiltonian).

| Intrinsic state | Energy [keV] | Structure of intrinsic excitations [individual components with percentage] | |
|---|---|---|---|
| gr. state $K=3/2^-$ | 21 | 3/2[761] (71%) <br> 3/2[642] +$Q_{30}$ (15%) | 3/2[501] (3%) <br> 3/2[631] +$Q_{30}$ (2%) |
| 1-st exc. state $K=3/2^+$ | 35 | 3/2[631] (37%) <br> 3/2[761] +$Q_{30}$ (17%) | 3/2[642] (29%) <br> 3/2[501] +$Q_{30}$ (7%) |
| 2-nd exc. state $K=5/2^-$ | 69 | 5/2[752] (53%) <br> 5/2[633] +$Q_{30}$ (42%) | |
| 3-rd exc. state $K=1/2^+$ | 77 | 1/2[631] (57%) <br> 1/2[761] +$Q_{30}$ (9%) | 1/2[640] (13%) <br> 1/2[770] +$Q_{30}$ (8%) |
| 4-th exc. state $K=1/2^+$ | 127 | 1/2[640] (47%) <br> 1/2[770] +$Q_{30}$ (13%) | 1/2[631] (19%) <br> 1/2[510] +$Q_{30}$ (9%) |

## Table 4

Comparison of theoretical and experimental values of the admixture parameter $\delta^2$(E2/M1)

| Transition | Exp. value | Theor. value $\eta_{cor}=0.50$, $\eta_{rec}=0.54$ | Theor. value $\eta_{cor}=0.59$, $\eta_{rec}=0.80$ |
|---|---|---|---|
| 9.2 keV → 0 keV | 0.695±0.248 | $1/2^+$(8.5 keV) → $3/2^+$(0 keV): 0.0008 | $3/2^+$(9.2 keV) → $1/2^+$(0 keV): 0.81 |
| 24.3 keV → 0 keV | 0.0116±0.0004 | $3/2^+$(39.2 keV) → $3/2^+$(0 keV): 0.0040 | $3/2^+$(40.4 keV) → $1/2^+$(0 keV): 0.0035 |
| 24.3 keV → 9.2 keV | 0.035±0.006 | $3/2^+$(39.2 keV) → $1/2^+$(8.5 keV): 0.0030 | $3/2^+$(40.4 keV) → $3/2^+$(9.2 keV): 0.0064 |

**Figure captions**

**Fig. 1** The β–decay scheme of $^{227}$Ac to $^{227}$Th as presented in Ref. [7].

**Fig. 2** An example of the $L_{1,2}$ and $M_{1-3}$ subshell conversion electron lines of the 24.3 keV and the 9.2 keV transitions, respectively, in $^{227}$Th (shown without correction for the spectrometer transmission dependence on the electron retarding voltage [13,14] and the $^{227}$Th half-life). The spectrum was measured at the absolute instrumental energy resolution of 14 eV and the energy step of 2 eV in two sweeps with the exposition time of 100 s per spectrum point in each sweep. The structure of the measured conversion electron spectra is complicated by a presence of the MNX group of Auger electrons of Th (indicated by the oblique lines in the picture).

**Fig. 3** An example of the $N_{1-3}$ and $L_3$ subshell conversion electron lines of the 9.2 keV and 24.3 keV transitions, respectively, in $^{227}$Th. The spectrometer was set to the 21 eV absolute instrumental energy resolution. The spectrum was scanned with the 2 eV step in four sweeps at the exposition time per spectrum point of 75 s in each sweep. The measured conversion electron lines are superimposed on the LMM Auger-electron spectrum of Th.

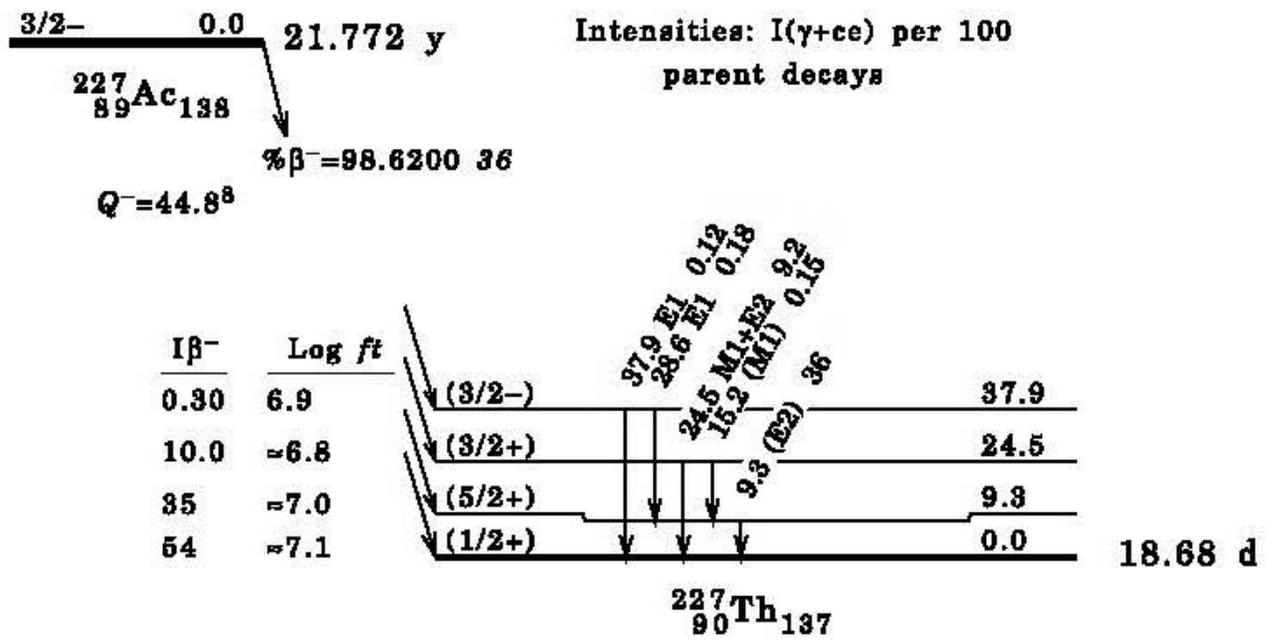

**Fig. 1** The β–decay scheme of $^{227}$Ac to $^{227}$Th as presented in Ref. [7].

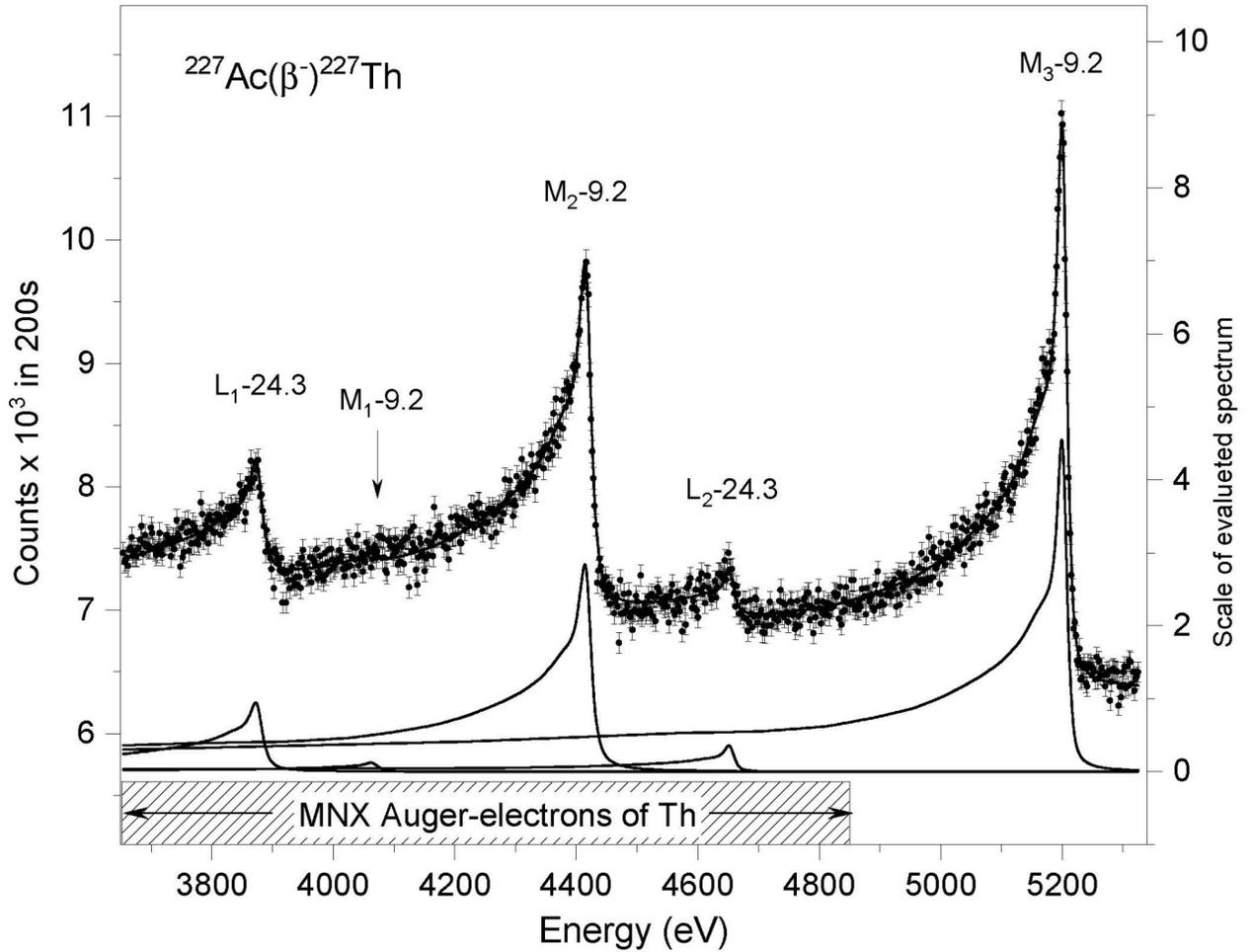

**Fig. 2** An example of the $L_{1,2}$ and $M_{1-3}$ subshell conversion electron lines of the 24.3 keV and the 9.2 keV transitions, respectively, in $^{227}$Th (shown without correction for the spectrometer transmission dependence on the electron retarding voltage [13,14] and the $^{227}$Th half-life). The spectrum was measured at the absolute instrumental energy resolution of 14 eV and the energy step of 2 eV in two sweeps with the exposition time of 100 s per spectrum point in each sweep. The structure of the measured conversion electron spectra is complicated by a presence of the MNX group of Auger electrons of Th (indicated by the oblique lines in the picture).

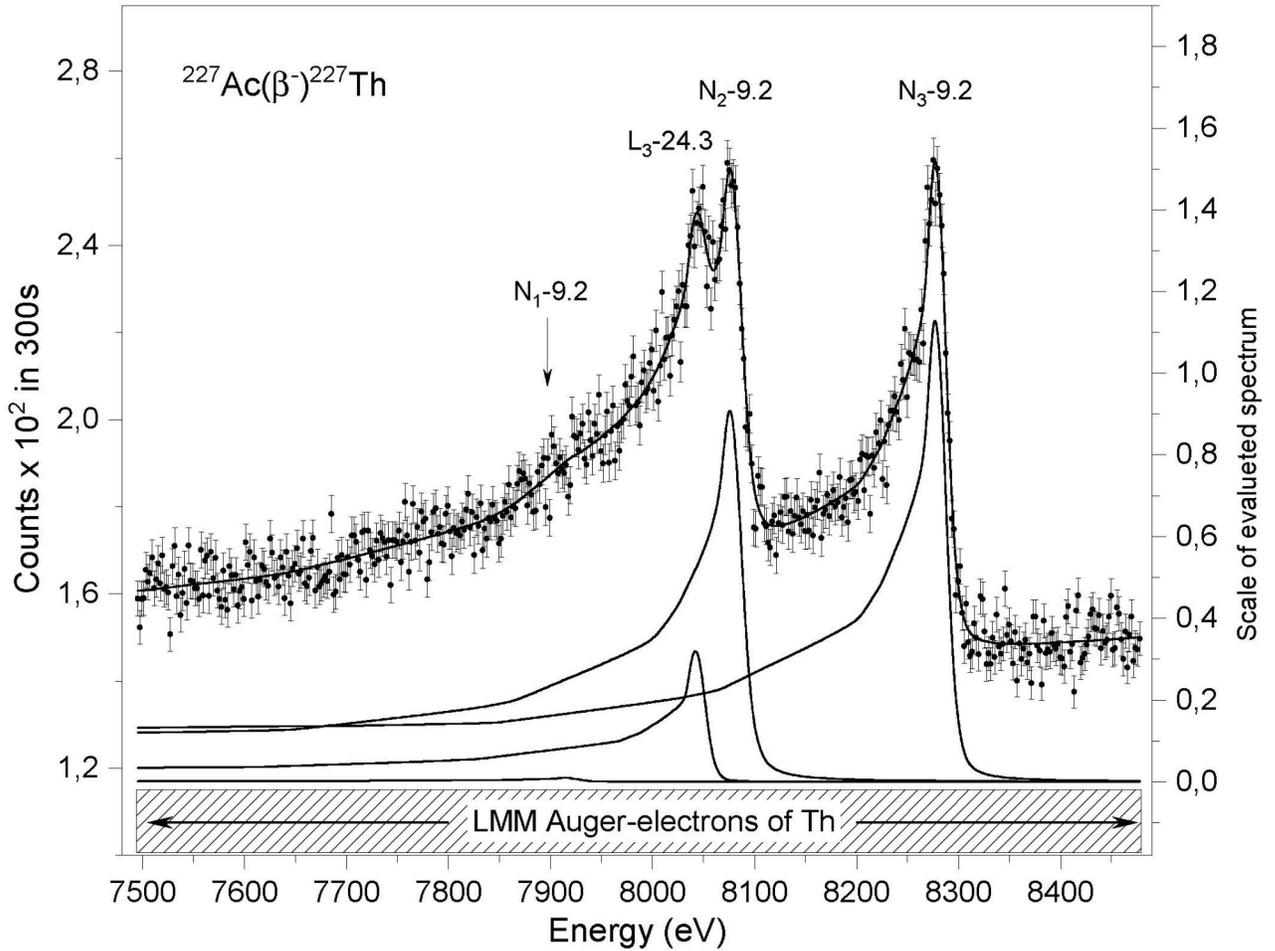

**Fig. 3** An example of the $N_{1-3}$ and $L_3$ subshell conversion electron lines of the 9.2 keV and 24.3 keV transitions, respectively, in $^{227}$Th. The spectrometer was set to the 21 eV absolute instrumental energy resolution. The spectrum was scanned with the 2 eV step in four sweeps at the exposition time per spectrum point of 75 s in each sweep. The measured conversion electron lines are superimposed on the LMM Auger-electron spectrum of Th.